# Lithium niobate waveguides with high-index contrast and preserved nonlinearity fabricated by High Vacuum Vapor-phase Proton Exchange

Alicia Petronela Rambu[1], Alin Marian Apetrei[1], Florent Doutre[2], Hervé Tronche[2], Vasile Tiron[3], Marc de Micheli[2] and Sorin Tascu[1],*

[1]*Research Center on Advanced Materials and Technologies, Science Department, Institute of Interdisciplinary Research, Alexandru Ioan Cuza University of Iasi, Blvd. Carol I, no. 11, 700506 Iasi, Romania*
[2]*Université Côte d'Azur, CNRS, Institut de Physique de Nice (INPHYNI), UMR 7010, Nice, France*
[3]*Research Department, Faculty of Physics, Alexandru Ioan Cuza University of Iasi, Blvd. Carol I, no. 11, 700506 Iasi, Romania*
*sorin.tascu@uaic.ro*

**Abstract:** Highly confining waveguides ($\Delta n_e$>0.1) without degraded nonlinear coefficient and low propagation losses have been fabricated in lithium niobate by a new process that we called High Vacuum Vapor-phase Proton Exchange (HiVac-VPE). Index contrast, index profile, nonlinearity and crystallographic phases are carefully investigated. Original analysis of index profiles indicates that the waveguides contains sub-layers whose depths are depending on the exchange durations. Propagation behavior, propagation losses and Second Harmonic Generation (SHG) response of HiVac-VPE channel waveguides are investigated at telecom wavelength. The results recommend HiVac-VPE as very promising technique for fabricating efficient nonlinear photonic integrated circuits in LN crystals.

## 1. Introduction

Due to its excellent electro-optical and nonlinear optical properties, paired with a relative easiness of fabricating waveguides, the lithium niobate (LN) crystal is an ideal platform for many integrated optics applications and photonic devices. Up to now, most of commercially available integrated optics devices or laboratory demonstrators are based on waveguides fabricated either by Ti-indiffusion [1,2] or by one of the conventional liquid-phase proton exchange techniques namely Annealed Proton Exchange (APE) [3-5], Reverse Proton Exchange (RPE) [6] or Soft Proton Exchange (SPE) [7-10]. All these techniques allow fabricating low loss waveguides with preserved electro-optical and nonlinear properties, but in return lead to low-index contrast typically in the range from $\Delta n_e$=0.01 to 0.03 [6,7,11]. These values limit the benefit linked to the confinement of the lights in a waveguide. It is important to note that the maximum index contrast allowed by the proton exchange technique is approximately $\Delta n_e$=0.12 at $\lambda$=633 nm [12], but most of waveguides exhibiting such a high-index contrast present a nonlinear coefficient $\chi^{(2)}$ totally degraded [13]. Two other techniques namely High Index Soft Proton Exchange (HISoPE) and High Vacuum Proton Exchange (HiVacPE) allowing the fabrication of channel waveguides that exhibit $\Delta n_e$=0.1 with almost unmodified nonlinearities have been recently reported [14,15]. These waveguides are most attractive for applications because they exhibit a high-index contrast which is adequate for tight light confinement allowing strong electro-optical and nonlinear optical effects. For now, the only downside of these two techniques is the strain induced propagation losses in the high-index contrast channel waveguides (>5dB/cm). However, given by a high control of the bath acidity due to the drying effect of the high vacuum at the beginning of the process, one of the biggest advantages of HiVacPE technique is the high stability and reproducibility of optical waveguide features [16]. Moreover, first integrated optics laboratory demonstrator fabricated by HiVacPE have already been investigated [17].

A very promising technique, namely Vapor-phase Proton Exchange (VPE), showing the possibility of creating high index contrast LN or lithium tantalate (LT) waveguides with preserved nonlinearity and low losses, has been reported in the late 1990s and the beginning of 2000s [18-26]. However, we lack some useful information about the waveguides fabricated by this technique such as: (i) the crystalline structure and the modification of lattice parameters of VPE planar waveguides are not precisely known due to the absence or a poorly chosen representation of XRD results; (ii) the nonlinearity, which is one of the greatest interests of LN waveguides, was investigated only for 40 nm in-deep layer under the surface of the waveguides [19,21,22] and deeper sub-layers have never been investigated; (iii) the investigations of modal behavior and propagation losses in channel waveguides are completely absent in literature. Finally, this technique has shown also some difficulties in controlling the depth of the waveguides [14,27,28]. As consequence, no further works and no functional optical device have been realized using VPE waveguides since the first aforementioned reports on this technique.

We previously demonstrated that the exchange process (in liquid bath) performed in a hermetically sealed hourglass tube at very low pressure, in order to diminish as much as possible any traces of water from the acid powders, leads to a new and very reproducible behavior in terms of index contrast, shape of index profile, nonlinearity, waveguide depth etc. [15,16].

Therefore, in this article, we present our work on the influence of the high vacuum drying effect on the crystallographic parameters, optical features and quality of both planar and channel waveguides fabricated in benzoic acid vapor. We will call this process HiVac-VPE for High Vacuum Vapor-phase Proton Exchange. Our approach has allowed to achieve very interesting results that have not yet been reported and also to compensate for the lack of information on the optical features of the fabricated waveguides by vapor phase method. The impact of high vacuum on the optical features of the waveguides and reproducibility of the process was tested over one year and a half by producing, in same vacuum conditions, more than fifty waveguides. In our study, both planar and channel waveguides have been fabricated and investigated. Over the time, no significant difference has been identified whether it is about index profile behavior, index contrast, nonlinearity or propagation losses. In addition, no aging phenomenon in the fabricated waveguides has been observed. We can say that HiVac-VPE is a trouble-free and highly reproducible process allowing producing efficient and compact devices for applications in modern photonics.

In section 2 we will detail the HiVac-VPE fabrication process. Section 3 and 4 will be devoted to the crystallographic and index profiles investigations respectively. In section 5 we will present an original and more precise analysis of index profiles and refractive index sub-layer depths of the protonated substrate. The nonlinearities on planar waveguides (which can be extrapolated to any waveguide shape) are presented in section 6. Section 7 will deal with the characterization of channel waveguides, essential for the fabrication of efficient photonic circuits. Finally we will present our conclusion and perspectives.

## 2. Fabrication of waveguides

High Vacuum Vapor-phase Proton Exchange (HiVac-PEV) process was performed on Z-cut LN samples cut from 3 inch optical grade wafers (supplied by Gooch & Housego) in a hermetically sealed hourglass tube for different exchange durations $t(h)$. Prior to be sealed, the bottom part of the tube was filled with 16g of pure Benzoic Acid (BA) powder (supplied by Sigma-Aldrich) as vapor source. The sample to be processed is placed in the top part of the tube and then, by using a turbo pumping station (HiCube 80 Eco equipped with a Compact Full Range Gauge PKR 251 – Pfeiffer), the tube is pumped down to a pressure as low as $p=3.4\times10^{-5}$ mbar. This very low pressure is $10^8$ times lower compared to the first reports on vapor phase exchange [18] and $10^5$ times lower compared to the ones reported by the literature for liquid phase exchange [14].

After the glass tube is sealed, it is placed into a metallic tube container ensuring a uniform heating and an easy and safe manipulation. The metallic tube is placed vertically in an oven preheated at 350°C which is the exchange temperature. As the temperature of the tube increases, the acid melts and boils at 122°C and 250°C respectively. After 30 minutes the acid vapor and the sample have reached the thermal equilibrium at the exchange temperature. From this moment we start counting the proton exchange duration $t(h)$. The bottleneck of the hourglass tube allows the sample to be separated from the liquid acid and to be exposed only to its vapor. The vapor pressure inside the glass is tube is constant and depends only on the exchange temperature as long as the acid is always in the liquid-vapor equilibrium at constant volume. Given the amount of acid, we can consider that the proton source is constant during the exchange process. At the end of the exchange duration the tube is cooled down to room temperature. The samples fabrication by HiVac-VPE process does not suppose any annealing or heat treatment after the vapor proton exchange process itself. Using this protocol, we fabricated planar and channel HiVac-VPE waveguides.

## 3. Crystallographic Characterization by X-Ray Diffraction

To investigate the crystalline structure and the modification of lattice parameters of HiVac-VPE planar waveguides, we used the X-ray diffraction method (XRD). We chose to use the rocking curves of crystallographic planes parallel to the surface of Z-cut LN samples with Miller indices: $h=0$, $k=0$, $l=12$ (00.12) corresponding to $\theta_B=41.807°$ Bragg angle in the substrate. This value is used as origin in the spectra reported on Fig. 1. It is worth to note that the lattice parameters of the proton exchanged layer are greater than those of the substrate [29]. In the case of a planar waveguide and considering the stress induced by the non-protonated substrate, the only allowed deformations are normal to the surface of the sample [30]. The strain $\varepsilon_{33}^{''}$ perpendicular to the surface can be obtained directly from the rocking curves. It is expressed as: $\varepsilon_{33}^{''} = -\Delta\theta_{hkl} \times \cot\theta_B$, where $\Delta\theta_{hkl}$ is the measured angular distance between the substrate peak and the protonated layer peak on the rocking curve from the surface plane $(hkl)$, in our case (00.12). It was demonstrated that the surface layer always presents the largest value of the strain [30]. Starting from the rocking curves we identify the crystallographic phases and also calculate the values of the strain and compare them to those reported in literature [20,21,25,30,31].

We present the rocking curves for virgin substrate and the samples protonated for *1h* and *5h* respectively. For the other samples the rocking curves shapes are located between these two samples. For an accurate comparison between our results and thus already presented in literature, we give the rocking curves in both logarithmic and linear scale.

As shown in Fig. 1, the HiVac-VPE waveguides present interesting crystallographic structures and exhibit characteristics that are worth to be noted. The *α*-phase is identified in all protonated samples and the maximum intensity peak of this phase is ranging from -133" in the sample protonated for *1h* to -162" in the sample protonated for *5h* respectively. Therefore, the *α*-phase peaks seat near of the one of the substrate and the crystallographic parameters vary gradually and remain similar to those of the substrate. This almost imperceptible shift (only 29") of *α*-phase peaks is easily visible only in logarithmic representation depicted in Fig. 1 (a). This fact was not presented in the previously reported works because the use of linear representations of the spectra like in Fig. 1 (b). Also, we clearly observe a sharp peak with its maximum at -474" in the sample protonated for *1h*. This peak corresponds to the $\kappa_2$-phase. For protonation during *5h* the $\kappa_2$-phase peak becomes more intense and shifts its maximum value to -496". The $\kappa_2$-phase peak intensity increase means a thickening of this layer at the surface of the sample. The bandwidth of the peak is close to the one of the substrate proving the high crystalline quality of $\kappa_2$-phase layer.

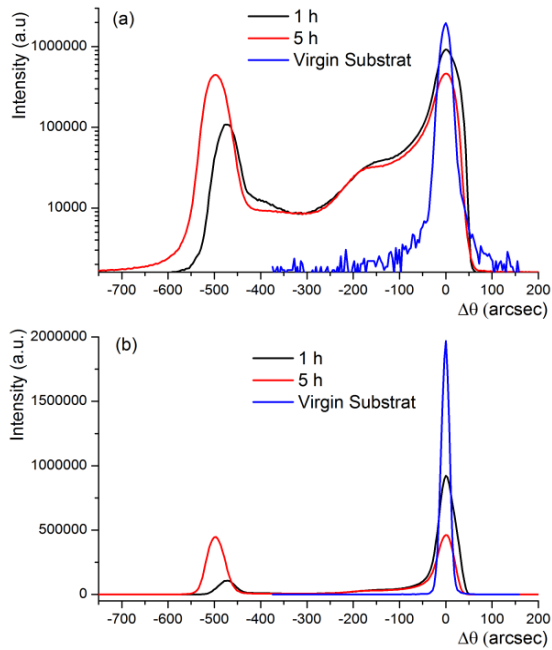

Fig. 1. X-ray rocking curves from (00.12) reflection of Z-cut HiVac-VPE planar waveguides represented in (a) logarithmic scale and (b) linear scale. The legends refer to the exchange duration *t(h)* at *T*=350°C.

It is worth to note that for all samples, the XRD signal does not go to zero between $\kappa_2$ and *α*-phase peaks. We clearly observe in log scale, as depicted in Fig. 1 (a), a large plateau around -350" which means that the transition between $\kappa_2$ and *α*-phase is an inhomogeneous layer with disordered crystalline structure where the crystallographic parameter varies gradually. This layer most likely contributes to relax the lattice parameters mismatch between $\kappa_2$ layer and *α*-phase layer and substrate. Note that for similar exchange durations and temperature conditions as in our experiments, previous works report the occurring of *β*-phase in waveguides fabricated by VPE in Z-cut LN [19]. In our experiments, the XRD study did not reveal any *β*-phase thanks to the drying effect of the high vacuum conditions. As previously demonstrated, diminishing as much as possible the water traces in the glass tube involves lower hydrogen and hydroxyl ion concentrations from dissociated water in the benzoic acid vapor [16]. As consequences, the acidity of the vapor in the glass tube of our experiments is lower than of the reported works.

In the following sections, we will see how the crystalline structure and the modification of lattice parameters are correlated to index profiles, nonlinearity and propagation losses exhibited by the waveguides.

## 4. M-lines measurements. Reconstruction of index profiles

In order to reconstruct the index profile of the planar waveguides whose crystalline structures have been presented, the effective indices of the propagation modes have been measured using a standard two-prisms coupling set-up and a He-Ne laser emitting at $\lambda$=633 nm [32]. The light was in-coupled and out-coupled using rutile (TiO$_2$) prisms pressed against the waveguide surface. At the output of the out-coupling prism, bright lines are observed and for each one the angle with the normal to the prism output surface is measured using an autocollimator (relative error on the angle is less than $10^{-5}$). This angle characterizes the propagation constant of the wave associated to the guided mode and thus we can determine the effective index $N_{eff}$ of each guided mode for a given waveguide. From the full-set of values of $N_{eff}$ of a given waveguide, one can reconstruct its index profile using the IWKB numerical method described in [33]. The measured effective indices $N_{eff}$ of TM guided modes at $\lambda$=633 nm as well as the raw values of mode depths and surface index calculated by IWKB allow us to reconstruct the raw index profiles of waveguides fabricated for different exchange durations by HiVac-VPE. As we can see on Fig. 2, it seems that the index profiles change from an exponentially decreasing one to a mix of a step and a gradient profile. The change from one regime to the other occurs between *t=2h* and *t=3h*. The exchange duration *t=3h* seems to be the threshold from which the index profile exhibits a step part which can be observed up to *t=5h* and beyond. Note that for *t=1h* and *t=2h* an abrupt jump is observed for the surface index.

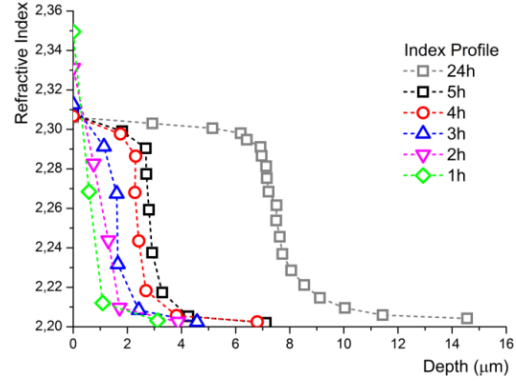

Fig. 2. Extraordinary refractive index profile at $\lambda$=633 nm reconstructed by IWKB for planar waveguides fabricated in Z-cut LN by HiVac-VPE at *350°C* with different exchange durations. The symbols represent the measured $N_{eff}$ of the propagating modes, except those on the ordinate that represent the raw surface indices calculated by IWKB. Dashed lines are guides for the eye.

It is worth to note the fact that propagation optical modes have also been identified in waveguides fabricated for exchange durations less than one hour, but the number of modes (one or two) at $\lambda$=633 nm makes no longer possible to use IWKB method.

This interpretation of the shape of index profiles and surface index value should be carefully considered because the IWKB method is inaccurate when the number of guided modes is small or an abrupt index change occurs in less than an optical wavelength. This drawback of IWKB method can be overcome and the index profiles can be corrected. Before proceeding to do this some facts have to be highlighted.

It is well known that the surface index of the waveguides fabricated by dipping the sample in acid melts depends only on the acidity of the proton source and not on the exchange duration (for long enough exchanges durations i.e. few hours). For the same acidity but different exchange durations the waveguide depth is the only difference between such fabricated waveguides. We assume that in the case of vapor-phase exchange the same thing happens. Therefore, as stated in the section II, once the liquid-vapor equilibrium is reached, the proton source is constant. So, for long enough HiVac-VPE process (i.e. *t=24h*) the fabricated waveguides exhibit the saturated value of surface index. Besides, the advantage of a very deep waveguide is that it supports many more optical modes compare to the waveguides fabricated for short exchange durations (i.e. *1h - 5h*). As a result, the IWKB method becomes more accurate when calculating the surface index value of a deep waveguide. In this way, the surface index value of the waveguide fabricated for *24h* exchange time will serve as maximum value that can be obtained by HiVac-VPE. For all others exchange durations (less than *24h*), the values of the surface index can be at most equal to the maximum value but never can exceed this. With these considerations we start to correct the index profiles.

The correction supposes the use of a similar approach that IWKB method but to impose the limit for the index surface value. So, starting from the experimentally measured effective indices $N_{eff}$, the correction supposes the modification of their calculated depths in order to obtain a surface index that can be at most equal to the maximum value (for *24h*) and never can exceed this. Note that we will consider only the mathematical solutions with physical meaning i.e. the value of corrected depth of $TM_n$ mode never can be greater than $TM_{n+1}$ mode and the value of index surface of the waveguide fabricated for the exchange duration *t* never can be greater than fabricated for the duration *t+1*. By doing so, the measured effective indices $N_{eff}$ of TM guided modes at $\lambda$=633 nm as well as the corrected mode depths and corrected surface index allow us to reconstruct the corrected index profiles of waveguides fabricated for different exchange durations by HiVac-VPE. Thus, the corrected index profiles of the waveguides fabricated for exchange durations in the range of *1h – 5h* as well as the IWKB raw index profile of the waveguide fabricated for *24h* are presented on Figure 3.

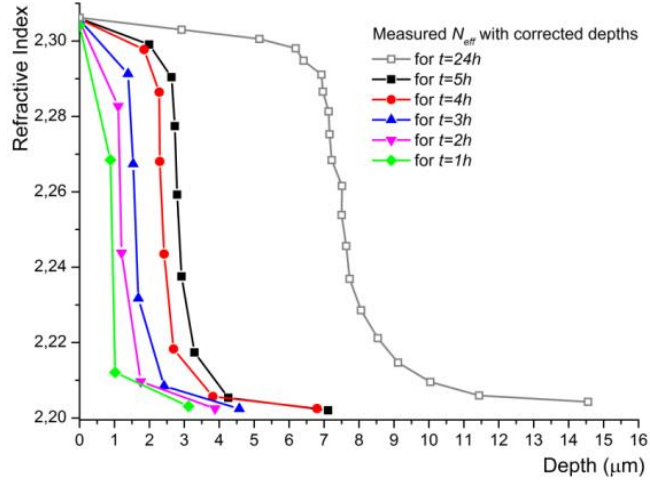

Fig. 3. Corrected refractive index profiles at $\lambda$=633 nm for planar waveguides fabricated in Z-cut LN by HiVac-VPE at 350°C for different durations. For $t$=24h is the raw index profile. The symbols represent the measured $N_{eff}$ of the propagating modes, except those on the ordinate that represent the corrected surface indices. Solid segment are guides for the eye.

On the Table 1 is presented the index contrast $\Delta n_e$ of a given waveguide calculated as the difference between its corrected surface index and the value of the extraordinary index of the substrate ($n_e$=2.2028 at 632.8 nm and room temperature).

Table 1. Index contrast $\Delta n_e$ at $\lambda$=633 nm of planar waveguides fabricated by HiVac-VPE process. All samples have been fabricated in Z-cut LN by using BA vapor at $T$=350°C.

| Exchange duration $t(h)$ | Corrected Surface Index | Index Contrast $\Delta n_e$ |
| --- | --- | --- |
| 1h | 2.3041 | 0.1016 |
| 2h | 2.3053 | 0.1028 |
| 3h | 2.3056 | 0.1031 |
| 4h | 2.3058 | 0.1033 |
| 5h | 2.306 | 0.1035 |
| 24h | 2.30621 | 0.10371 |

In the next section a thorough analyze of index profiles will be done by using the mathematical method fully described in previously reported work [15].

## 5. Analysis of Index Profiles. Layer depths and index profiles features

In this section, by using the method fully described in [15], we thoroughly analyze the corrected index profiles for *1h-5h* exchange durations and IWKB index profile for *24h* respectively. Thus, it one determines the best fit for the experimental points depicted in Fig. 3 and extracting the waveguides features. However, the best fit is obtained when a sum of two generalized exponential functions is used. It is expressed by:

$$n(d) = n_e + A_1 \exp\left(-(d/w_1)^{a_1}\right) + A_2 \exp\left(-(d/w_2)^{a_2}\right) \tag{1}$$

where, $A_1$, $A_2$, $w_1$, $w_2$, $a_1$ and $a_2$ are adjustable parameters and take different values depending on exchange time and $d$ is the depth of the waveguide. Each generalized exponential function in (1) describes the behavior of one of the chunk (step and gradient) of the index profiles [15]. On Fig. 4 are plotted index profiles for the Z-cut samples fabricated by HiVac-VPE. The symbols represent the measured $N_{eff}$ of the propagating modes, except the corrected surface indices on the ordinate. The solid lines are the fits obtained by using Eq. (1).

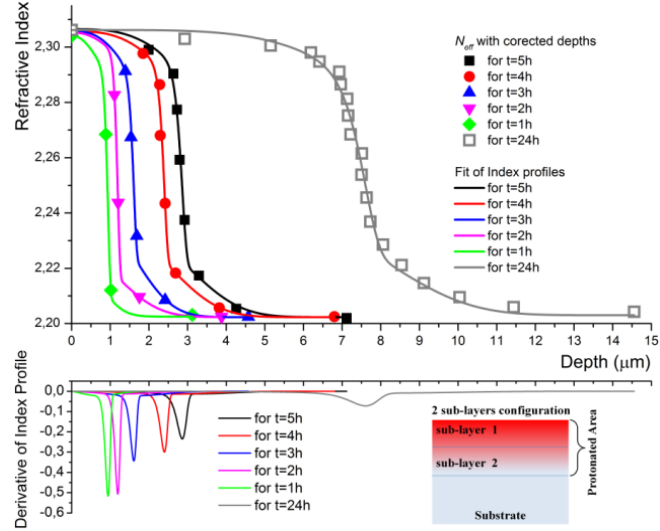

Fig. 4. Top: index profiles of Z-cut HiVacPE waveguides fabricated for different exchange durations. The symbols represent the measured $N_{eff}$ of the propagating modes, except the IWKB corrected surface indices on the ordinate. The solid lines are the fits obtained by using Eq. (1). Bottom: derivative of the fits. Inset: sub-layers structures of the waveguides. The intensity of the red color suggests the refractive index value in the waveguides.

As it was presented in [15], useful information can be obtained from derivatives of index profiles curves. As we clearly can see in the bottom part on Fig. 4, the derivatives of the fits of index profiles exhibit systematically one minimum point for all fabricated waveguides. These minimum points correspond to a given depth in the waveguide where, there is a significant refractive index difference between two layers. Only one minimum point means that the waveguide exhibits two refractive index sub-layers. First sub-layer starts from the surface of the waveguide and extends, as expected, to an increasing depth depending on the exchange duration until approximately 0.94 μm, 1.19 μm, 1.6 μm, 2.4 μm, 2.86 μm and 7.6 μm in depth. Then the second sub-layer starts and spreads to the end of the protonated area (see inset in bottom part on Fig. 4 for two sub-layers configuration of the waveguide).

Note that HiVac-VPE allows to pass the index contrast at $\Delta n_e \geq 0.1$ at shorter exchanges than what we could find in literature with similar exchange temperature (for example, $\Delta n_e = 0.1016$ for $t=1h$ (see Table 2) versus $\Delta n_e < 0.01$ in [19]). There is a factor ten of difference between the index contrast values. Even for long exchange durations $t \geq 3h$ the maximum values for index contrast reported by the literature is less than 0.1 at $\lambda = 633$ nm [19,20].

We recall once again that all our results presented above show a very high reproducibility (no significant differences over fifty waveguides processed over one year and a half) never reached before, thanks to the drying effect of high vacuum pumping.

## 6. Nonlinearity measurements

One of the great interests of LN waveguides is their high nonlinear efficiency. As we stated in the introductory section, the information about the nonlinearity of VPE waveguides is limited [19,21,22]. In these studies, the incident light beam at 532 nm was directed onto a planar waveguide surface and the reflected second harmonic signal at 266 nm was measured. This method is fully described in reference [13]. Since the wavelength of the second harmonic is strongly absorbed by the waveguide and substrate lattice (absorption edge at 330 nm), only the harmonic generated in the first 40 nm in-deep layer under waveguides surface goes out of the sample and could be detected. Consequently, this method, by unfortunate choice of fundamental and second harmonic wavelengths, eliminates SHG contribution of dipper sub-layers of the waveguide.

In order to overcome the material absorption at UV wavelengths, the nonlinearity of our HiVac-VPE waveguides was probed by a method similar to the surface SHG microscopy [34]. Our experimental setup is fully described in [35]. A laser beam at wavelength 1550 nm is focused on the polished end facet of a waveguide and vertically scanned through the different regions: air, protonated layers and substrate respectively. By measuring, during the scan, the reflected fundamental and second harmonic (SHG) beams, one can qualitatively evaluate the variation of the second–order nonlinearity coefficient induced by the waveguide fabrication process. The fundamental beam is used to identify the air-waveguide interface position and the intensity of the harmonic is used to compare nonlinearities of protonated layer and substrate respectively.

In Fig. 5 are presented typical examples of these scans superimposed with the corresponding index profile at $\lambda = 633$ nm for a Z-cut waveguide fabricated for the exchange duration *1h* (a) and *5h* (b) respectively.

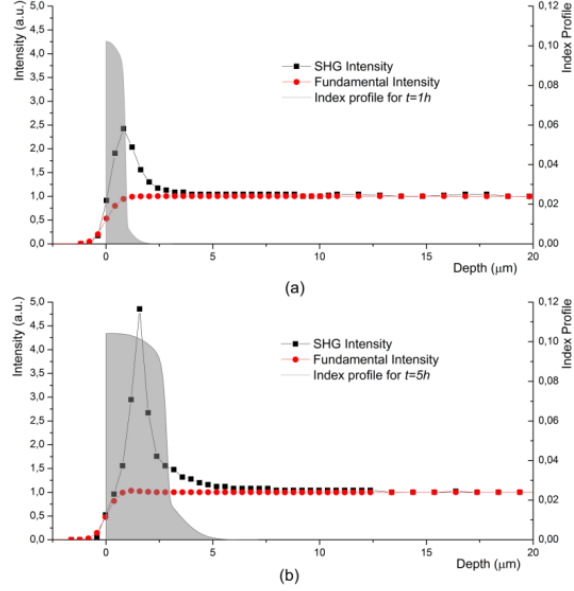

Fig. 5. SHG profiles and reflected fundamental signal of Z-cut HiVac-VPE waveguides superimposed with index profiles (region in gray color) for (a) *t=1h* and (b) *t=5h*.

The SHG signal superimposed on the refractive index profile of the waveguide clearly shows an intense peak of SHG in the protonated area. This enhancement of the SHG signal is probably a complex function of the discontinuity of the nonlinear coefficient $d_{33}$ between the air and the crystal [36] and of the confinement of the pump beam due to the high index contrast in the step part (first sub-layer) of the waveguide. Therefore, without a complete understanding of the enhancement of the reflected SHG signal and due to the lateral resolution of the experiment (~2μm) and the small depth of the $\kappa_2$ layer (0.94μm to 2.84μm), it is impossible to extract information about the exact value $d_{33}$ in HiVac-VPE waveguides. However, we can assume that the nonlinear coefficient value of the waveguide is at least equal to the substrate value, so there is no degradation of the nonlinear coefficient $d_{33}$.

Therefore, by our method, we have shown that nonlinearity, essential for efficient nonlinear applications and devices, are conserved for whole protonated layer of the waveguides fabricated by HiVac-VPE. In addition, no aging phenomenon in the fabricated waveguides has been observed.

## 7. Investigation of channel waveguides

As we mentioned in the introductory section, investigations on channel waveguides fabricated in benzoic acid vapor are completely missing on the literature even if they are paramount elements of modern photonic circuits. In this section, we propose a thorough analyze of HiVac-VPE channel waveguides in order to broaden the knowledge about this kind of structures. In view of integrated optics applications, we have targeted the fabrication of channel waveguides single-mode at *λ*=1550 nm. Taking into account the high index contrast, we have to adjust the width and the depth of the waveguides so that they remain single-mode at the wavelength of interest. By using the fabrication procedure described in section II, the channel waveguides were made through $SiO_2$ mask exhibiting openings of 1 μm, 1.5 μm and 2 μm widths respectively. The depth of the waveguides is controlled by the exchange duration, so *t=1h* in this case. The mask was deposited on the surface of Z-cut sample so the waveguides will support TM modes only. Besides the channel waveguides, the so prepared samples have the advantage to exhibit on the opposite surface, a planar waveguide very useful for preliminary characterizations described in previous sections. Given the narrow openings on the mask it is worth to note that: (i) the channel waveguides should exhibit smaller depths than the planar waveguide as the kinetics of proton exchange is slower through narrow openings; (ii) the channel waveguides will probably exhibit larger widths than the mask as the lateral diffusion of protons is no longer negligible through narrow openings [37,38].

To allow characterizing the channel waveguides, after the exchange process, the input and the output edges of the samples were polished at optical grade quality. The length of the waveguides after polishing is 9.77±0.01mm. The modal behavior was verified by observing the near-field pattern of the propagating mode for each width of the fabricated channel waveguides. The investigations were performed by using a tunable laser (Yenista Optics – Tunics T100R) around *λ*=1550 nm. For the in-coupling we used a micro lensed polarization maintaining fiber. The image of the propagating modes at the waveguide output is obtained with a microscope objective and an infrared highly sensitive CCD camera (NIR-300PGE from VDS Vosskühler GmbH).

In the case of waveguides with both 1 μm and 1.5 μm width respectively, we experimentally validated a single-mode propagation behavior at *λ*=1550 nm. In Fig. 6 we present an imaging of fundamental modes, as well as mode intensity distribution in width and depth. Fig. 6 (a) shows the result for waveguides fabricated through

1μm silica mask opening. The FWHM measurements of intensity profile are 2.30±0.15μm in width and 2.31±0.15μm in depth. Fig. 6 (b) presents the result for waveguides fabricated through 1.5μm silica mask opening. The FWHM measurements of intensity profile are 2.80±0.15 μm in width and 2.92±0.15μm in depth.

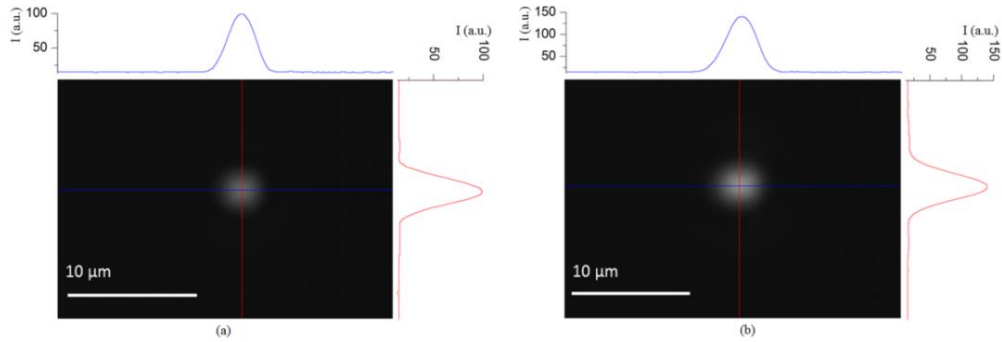

Fig. 6. Near-field imaging of the modes at the output of channel waveguides fabricated at *T=350°C* for *1h* through silica mask with opening of (a) 1 μm and (b) 1.5 μm width respectively.

In Fig. 7 we present the near-field pattern at *λ*=1550 nm of the optical modes supported by a channel waveguide with 2 μm width. As we clearly seen, these waveguides exhibit a slightly multimodal propagation. In Fig 7 (a) is an image of the fundamental mode supported by such waveguide and in Fig 6 (b) it is seen excited both the fundamental mode and the first superior mode.

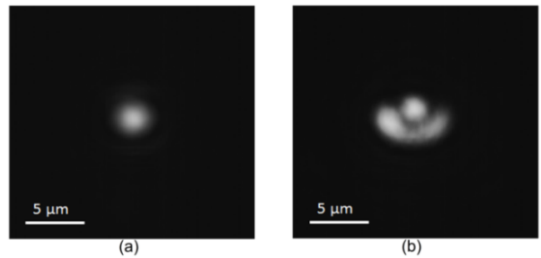

Fig. 7. Near-field imaging of the modes at the output of channel waveguides fabricated at *T=350°C* for *1h* through silica mask with opening of 2 μm width. (a) fundamental mode and (b) superior mode + fundamental.

These were the only modes experimentally observed. So, for 2 μm, the waveguides exhibit a multimode behavior because of the high index contrast, even for exchange short durations of exchange i.e. for one hour. Is obvious that the propagation behavior in HiVac-VPE channel waveguides will be largely multimode at *λ*=1550 nm for any other combination of waveguide width and exchange duration greater than the investigated ones.

Propagation losses in the HiVac-VPE channel waveguides have been measured by using the Fabry-Pérot cavity technique [39]. Because this technique requires single-mode propagation through the waveguides, the measurements were performed only for the waveguides exhibiting 1.5 and 1 micron width respectively. The best of our investigated channel waveguides exhibit propagation losses around 3.5±0.1 dB/cm.

As the planar waveguides on the opposite face of the samples exhibited quite good optical quality during M-lines investigations, not compatible with measured high propagation losses, we suppose the arising of hybrid modes (EH modes) in channel waveguides. Indeed, the complex strain tensor induced by protonation causes TM-TE polarization coupling in the channel waveguides leading to arise of hybrid modes (EH modes) whose ordinary part radiates in the substrate inducing propagation losses.

This kind of modes were already observed in proton exchanged Z-cut LN channel waveguides and definitely related to the important strain induced by the proton exchange process [14,15,40,41]. To check this hypothesis we observed the far field at the output of the channel waveguide by using a visible light at *λ*=633 nm. In order to separate TM and TE polarizations, a polarizer was introduced between the sample and the screen placed just a few centimeters away. When only TE polarization is passing, it can be observed on images in Fig. 8 the characteristic semicircles corresponding to the light radiating into the substrate, the channel waveguide acting like an antenna. A part of radiating light is reflected by the bottom surface of the sample, resulting in mirrored semicircles in the upper half of the images on Fig. 8.

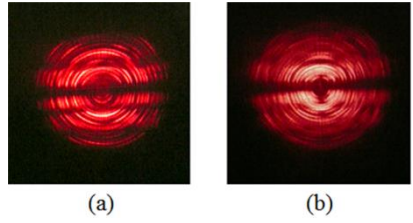

Fig. 8. Far field picture on a screen of TE polarized hybrid modes in HiVacPE channel waveguide at λ=633 nm for (a) 1 μm width and (b) 1.5 μm width respectively.

These EH modes are not observed in planar Z-cut LN waveguides during M-lines measurements because in this case the strains and stresses induced by the exchange process do not modify the symmetry of the crystal and no TE-TM coupling occurs. Contrariwise, in the channel waveguides, where the strains and stresses are induced by the presence of the non-modified crystal under and on the sides of the waveguide this one becomes biaxial, thereby generating EH modes. Therefore, is clear that the mechanism responsible for propagation losses in channel HiVac-VPE waveguides is the occurrence of hybrid modes caused by TM-TE polarization coupling. Note that two scenarios may appear when talking about EH modes. First, if the effective index $N_{eff}$ of a EH mode is higher than ordinary index $n_o$ of LN it propagates like a pure TM mode, and its ordinary component has an insignificant role. Second, if $N_{eff}$ of a EH mode is lower than $n_o$ of LN then, the ordinary component of the mode radiates into the substrate. As we experimentally validated the second one has been occurred in our case.

## 8. Discussions and Conclusions

In this work, we took advantage of the fact that, using the drying effect of the high vacuum at the beginning of the fabrication process, the waveguides fabricated by vapor proton exchange at *350°C* simultaneously exhibit strong confinement, preserved nonlinearities and most important high reproducibility of their optical features and geometrical parameters. If the main purpose was to improve the reproducibility and the quality of the produced waveguides by limiting and controlling the water traces in the glass tube, we discovered that the high vacuum involves new behaviors and results. The impact of high vacuum on the fabrication reproducibility was tested by producing in similar conditions more than fifty waveguides during one year and a half and no notable differences were identified. In addition, no aging phenomenon in the fabricated waveguides has been observed for a year.

From crystallographic point of view all investigated samples exhibit only *α* and *κ₂* crystallographic phases.

Regarding the optical features of the HiVac-VPE waveguides we discovered very interesting behaviors. For example, the high index contrast $\Delta n_e \geq 0.1$ are obtained for relatively short exchange durations (*t=1h*). The index profile is a mix of a step and a gradient for all investigated samples. From derivatives of the index profiles fits, we showed that the HiVac-VPE waveguides exhibit two refractive index sub-layers, with the depth of the first one depending on the exchange duration.

Concerning the nonlinearity of HiVac-VPE waveguides, they were probed for the whole in-deep protonated layers, leading to the conclusion that the nonlinearities are at least equal to the one of the substrate.

Finally, for the first time as far as we know, we investigated the modal behavior and propagation losses of channel waveguides fabricated in acid vapor at telecom wavelength. Single-mode propagations at telecom wavelengths were obtained for channel waveguides exhibiting 1 and 1.5 μm width and protonated for *1h*. Propagation losses (3.5±0.1 dB/cm) have been measured by using the Fabry-Pérot cavity technique. We showed that these relatively high propagation losses were due to hybrid modes caused by TM-TE polarization coupling in channel waveguides. Future work should be to decrease the propagation losses measured in this work and we introduce here some leads to do so: (i) the fabrication of a planar HiVac-VPE waveguide does not induce lateral stress, therefore a channel waveguide fabricated afterward using a surface etching of the sample to define ridge channel waveguides should eliminate the arising of EH modes; (ii) decrease the ordinary index $n_o$ of the substrate prior to the HiVac-VPE waveguide fabrication. This could be obtained with a deep planar waveguide which exhibit low-index variation This would not significantly affects the afterward afterwards the high HiVac-VPE index contrast, while the reduction of $n_o$ should be enough to have the effective index $N_{eff}$ of the fundamental propagating mode in the channel waveguide larger than the ordinary index of the modified substrate, thus preventing from propagation losses due to radiation.

In summary, we have demonstrated that intensively pumping the exchange ampoule to diminish as much as possible the water content in the exchange glass tube resulted in a new proton exchange technique the so-called High Vacuum Vapor-phase Proton Exchange (HiVac-VPE). Taking advantage of simultaneously exhibited index contrast as high as $\Delta n_e=0.1$, relatively low propagation losses and perfect preserved nonlinearity, the HiVac-VPE technique can be a way of improving the efficiency of the nonlinear components and devices in many applications using LN waveguides. Besides, channel waveguides exhibiting high-index contrast with preserved non-linearity allow fabricating nonlinear photonic wires. This is of high interest for nonlinear highly efficient and compact devices for quantum and classical optical data processing.